\documentstyle[11pt,fleqn]{article}
\def\beeq{\begin{equation}}
\def\eneq{\end{equation}}
\def\beeqa{\begin{eqnarray}}
\def\eneqa{\end{eqnarray}}
\def\nel{N_{\rm el}}
\def\nc{N_{\rm c}}
\def\nb{N_{\rm b}}
\def\tc{T_{\rm c}}
\def\soc{{\rm C}_{60}}
\def\rug{{\rm C}_{70}}

\def\eps{\varepsilon}

\addtocounter{section}{-1}
\begin{document}

\begin{center}

{\large{\bf{Lattice distortion and energy level structures \\
in doped C$_{60}$ and C$_{70}$ molecules \\
studied with the extended Su-Schrieffer-Heeger model: \\
Polaron excitations and optical absorption
} } }

\vspace{1cm}

{\rm Kikuo Harigaya$^*$}\\

\vspace{1cm}

{\sl Fundamental Physics Section, Physical Science Division,\\
Electrotechnical Laboratory,\\
Umezono 1-1-4, Tsukuba, Ibaraki 305, Japan}

\vspace{1cm}

(Received 5 November 1991)
\end{center}

%\maketitle

\Roman{table}
%\large

\vspace{1cm}

\noindent
{\bf ABSTRACT}

\noindent
We extend the Su-Schrieffer-Heeger model of polyacetylene to
$\soc$ and $\rug$ molecules, and solve numerically.
The calculations of the undoped systems
agree well with the known results.  When the system ($\soc$ or $\rug$)
is doped with one or two
electrons (or holes), the additional charges accumulate almost
along an equatorial line of the molecule.
The dimerization becomes weaker almost along the same line.
Two energy levels intrude largely in the gap.  The intrusion is larger
in $\rug$ than in $\soc$.
Therefore, ``polarons'' are predicted in doped
buckminsterfullerenes.  We calculate optical absorption coefficient
for $\soc$ in order to look at how ``polarons'' will be
observed.  It is predicted that there appears a new peak at the
lower energy than the intergap transition peaks.
It is also found that $\soc$ and $\rug$ are related mutually with
respect to electronical structures
as well as lattice geometries.

{}~


\pagebreak

%%%%%%%%%%%%%%%%%%%%%%%%%%%%%% input document
% %%%%%%%%%%%%%%%%%%%%%%%%%%%%%%%%%%

\section{INTRODUCTION}

Recently, the ``buckminsterfullerenes" C$_N$ have been intensively
investigated.  Particularly, the occurrence of the
superconductivity in C$_{60}$ crystals doped with alkali metals
($\tc = 18$K for K$^1$, $\tc = 28$K for Rb$^2$, and $\tc = 33$K
for Rb and Cs$^3$) has attracted much interests in electronic
structures of undoped and doped systems.

Band calculations$^4$ of the undoped C$_{60}$ crystals have predicted the
large gap ($\sim$ 2eV) like an insulator.  The bands have weak dispersions and
their structures reflect molecular orbitals of C$_{60}$.$^5$
When $\soc$ is doped with alkali metal ions, it is expected that the
Fermi level is located in weakly dispersive bands.  It has been proposed
that electrons around the Fermi level condense into a superconducting
ground state by attractive interactions which are mediated particularly by
phonons.$^{6-9}$  The calculation of C$_{70}$ by
the local-density approximation$^{10}$ and quantum
chemical calculations$^{11}$
has also shown the presence of the large gap.

There are several evidences that electronic structures change when
buckminsterfullerenes are doped.  For example,
photoemmission studies$^{12}$ of C$_{60}$ and C$_{70}$ doped with alkali
metals have shown appearance and shift of peak
structures, which cannot be described by a simple band-filling picture.
The changes in C$_{60}$ are different from those in C$_{70}$.

The electron spin resonance (ESR) study on the radical anion
of C$_{60}$ by Kato et al.$^{13}$
has revealed the small $g$-factor, $g=1.9991$,
and this is associated with the residual orbital angular momentum.
The irregular vibrational structure of the electronic absorption
has been suggested to be due to the Jahn-Teller distortion.
Therefore, it is again revealed that lattice and electronic structures
change in doped C$_{60}$.

In the present paper, we concentrate
upon a single C$_{60}$ or $\rug$ molecule, and
investigate lattice distortion and reconstruction of electronic levels
upon doping.   It is much possible that electronic states are spread over
the surface of the molecule.  Therefore, correlation effects might
be neglected in the first approximation.  We assume that the effects are
renormalized into effective one electron levels.

First, we describe C$_{60}$ as an electron-phonon system and
extend the Su-Schrieffer-Heeger (SSH) model$^{14}$ which was applied
successfully for {\sl trans}-polyacetylene.
We find that length difference between the short and long bonds
becomes smaller along the equatorial line of the sphere when the change in the
electron number is one and two (the ``weakly-doped'' case).
Energy level structures also change.
One energy level splits upward from the fivefold degenerate highest occupied
molecular orbital (HOMO), while the other energy level splits downward from
the threefold degenerate lowest unoccupied molecular orbital (LUMO).
The changes are like those in the polaron and bipolaron excitations
in polyacetylene and undegenerate conjugated polymers.$^{15}$
The reduction in the degeneracies of wave functions is consistent with
that by the Jahn-Teller theorem.  Therefore, the symmetry breaking described
by the extended SSH model would be one of
the realizations of Jahn-Teller distortions.
We suggest that these changes might be the origin of the experimental
findings on the changes of electronic structures.$^{12,13}$
We calculate optical absorption coefficient in order to look at
how our remarkable change in energy level structures
will be observed.  It is found that a new
absorption peak appears at much lower frequency than the intergap
transition peaks of the undoped $\soc$.  Peak positions are mutually different
between electron and hole dopings.  Our calculations are extended
to ``heavily-doped'' systems, too.  Electron number is changed between
the maximum and minimum numbers, which can be realized in principle.
As the doping proceeds, the average dimerization becomes more and more
weaker.  When six electrons are doped, the average is about 0.03 times of that
of the undoped $\soc$.  This value would be hardly observed in experiments.
When ten holes are doped, the average is about
0.5 times.  In this case, the dimerization would remain more probably
than in the electron-doped case.

Secondly, we investigate possible changes
of lattice and electronic systems in the $\rug$ molecule, which
is present abundantly as well as $\soc$.  Even though the spatial symmetry
of $\rug$ is lower than that of $\soc$, we certainly
expect that the interesting
changes might occur as in $\soc$.  We show that the extended SSH model
well describes the lattice and energy level structures of the undoped $\rug$,
comparing with the other results.$^{10,11}$
Particularly, the same parameters as
in $\soc$ are valid.  Then, we calculate for systems
where one or two electrons are added
or removed.  It is predicted that the HOMO and LUMO of
the undoped system intrude largely into the
energy gap.  The intrusion is larger than in $\soc$.
The lattice pattern deforms from that
of the undoped system.  We first find that sites, where additional charges
favor to accumulate, are common to $\soc$ and $\rug$.  They are
along the equatorial line in $\soc$.  Wave functions
of the intragap levels have large amplitudes at corresponding sites.
This should be compared to the statement that $\rug$ is made from
$\soc$, by cutting into two parts
and adding ten carbons.  It is quite interesting
that there are relations of electronical properties as well as the
structural relation, between (doped as well as undoped)
$\soc$ and $\rug$.

We note that the present paper is a detailed report of two short
publications$^{16,17}$ where partial data are reported and discussed.

This paper is organized as follows.
In Sec. II, the model and the numerical method are explained.  In Sec. III,
results of $\soc$ are shown and discussed.  In Sec. IV, we show results of
$\rug$.  We close the paper with several remarks in Sec. V.

\section{MODEL AND NUMERICAL METHOD}

The SSH model$^{14}$ is extended for the C$_{60}$ and $\rug$ molecules,
\beeq
H = \sum_{\langle i j \rangle,s} [- t_0 - \alpha (u_i^{(j)} + u_j^{(i)})]
(c_{i,s}^\dagger c_{j,s} + {\rm h.c.})
+ \frac{K}{2} \sum_{\langle i j \rangle} (u_i^{(j)} + u_j^{(i)})^2.
\eneq
In the first term, the quantity $t_0$ is the hopping integral of the
undimerized systems; $\alpha$ is the electron-phonon coupling; the
operator $c_{i,s}$ annihilates a $\pi$-electron at the $i$-th carbon atom
with the spin $s$; $u_i^{(j)}$ is the displacement of the $i$-th atom
in the direction of the $j$-th atom (three $u_i^{(j)}$ for the given $i$
are mutually independent); the sum is taken over nearest neighbor
pairs $\langle i j \rangle$.  The quantity, $u_i^{(j)} + u_j^{(i)}$,
is the change of length of the bond between the $i$- and $j$-th atoms.
When it is positive, the hopping integral increases from $t_0$;
accordingly we take the sign before $\alpha$ to be negative.
The second term is the elastic energy of the
phonon system; the quantity $K$ is the spring constant.

The model Eq. (2.1) is solved by the adiabatic approximation for phonons.  The
Schr\"{o}dinger equation for $\pi$-electron is
\beeq
\eps_\kappa \phi_{\kappa,s} (i) = \sum_{\langle i j \rangle}
(-t_0 - \alpha y_{i,j} ) \phi_{\kappa,s} (j),
\eneq
where $\eps_\kappa$ is the eigenvalue of the $\kappa$-th eigenstate
and $y_{i,j} = u_i^{(j)} + u_j^{(i)}$ is the bond variable.  The
self-consistency equation for the lattice is
\beeq
y_{i,j} = \frac{2\alpha}{K} {\sum_{\kappa,s}}^{'} \phi_{\kappa,s}(i)
\phi_{\kappa,s}(j)
- \frac{2\alpha}{K} \frac{1}{N_{\rm b}} \sum_{\langle k l \rangle}
{\sum_{\kappa,s}}^{'} \phi_{\kappa,s}(k) \phi_{\kappa,s}(l),
\eneq
where the prime means the sum over the occupied states, the last term is
due to the constraint $\sum_{\langle i j \rangle} y_{i,j} =0$, and
$N_{\rm b}$ is the number of $\pi$-bonds.  We use $\nb = 90$ for $\soc$
and $\nb = 105$ for $\rug$.  The same constraint has
been used in polyacetylene$^{18}$ in order to numerically obtain a correct
dimerized ground state of the undoped system.  We can certainly assume that
the constraint works well for the present systems, too.  Owing to the
constraint, we can avoid the contraction of the lattice; the total
energy width from the lowest occupied molecular orbital to the highest
unoccupied molecular orbital does not vary from that of the undimerized
system.

A numerical solution is obtained in the following way:
\newline
i) Random numbers between $-y_0$ and $y_0$ ($y_0 = 0.1{\rm \AA}$)
are generated for the initial values of the bond variables
$\{y_{i,j}^{(0)}\}$.  Then, we start the iteration.
\newline
ii) At the $k$-th step of the iteration, the electronic part of the
hamiltonian is diagonalized by solving Eq. (2.2) for the set of the
bond variables $\{y_{i,j}^{(k)}\}$.
\newline
iii) Using the electronic wave functions $\{\phi_{\kappa,s}(i)\}$ obtained
above, we calculate the next set $\{y_{i,j}^{(k+1)}\}$ from the left hand side
of Eq. (2.3).
\newline
iv) The iteration is repeated until the sum
$\sum_{\langle i,j \rangle} [y_{i,j}^{(k+1)}-y_{i,j}^{(k)}]^2$
becomes negligibly small.
\newline
v) It is checked that there is only one stationary solution for each
electron number by changing the initial random set $\{y_{i,j}^{(0)}\}$
and the maximum value $y_0$.  Here, we regard
that possible degenerate solutions with respect
to the geometrical direction of the soccer or rugby ball are an identical
solution.

\section{UNDOPED AND DOPED C$_{60}$ MOLECULES}

We take $t_0 = 2.5$eV, because a tight binding model with the same value
well reproduces the band structure of a two-dimensional graphite plane.$^{19}$
The same value is used in polyacetylene.$^{14}$
This indicates that the origin of $t_0$ is common.  It is quite interesting
to point out that the same parameter $t_0$ is valid for three different
systems: graphite, polyacetylene, and C$_N$.  Therefore, the overlap
of $\pi$-orbitals of nearest-neighbor carbon atoms is almost independent
of kinds of materials, even though bond lengths and bonding angles are mutually
different.

Two quantities, $\alpha$ and $K$, are determined so that
the length difference between the short and long bonds is the
experimentally observed value: 0.05\AA.$^{20}$
Here, the dimensionless electron
phonon coupling $\lambda \equiv 2\alpha^2 / \pi K t_0$ is taken as 0.2 as
in polyacetylene.$^{14}$  The choice of $\lambda$ stands upon the analogy that
the general parameters of $\pi$-conjugated carbon systems do not vary
so widely. The result is $\alpha = 6.31$eV/\AA{\ } and $K=49.7$eV/\AA$^2$.
This choice of the parameters are valid enough,
in view of the point that the resulting
energy gap of the undoped system 2.255eV agrees with about 1.9eV of
the other result$^5$ within 20\%.  The sensitivity in the choice of parameters
does not depend on the qualitative results of this paper, for example,
intrusion of levels, lattice distortion patterns, and so on.

Number of electrons $N_{\rm el}$ is varied within $N-10 \leq N_{\rm el}
\leq N+6$, where $N(=60)$ is the number of carbon atoms.  When
$N_{\rm c} \equiv N_{\rm el} - N$ is positive, the system is doped with
electrons.  If $N_{\rm c} < 0$, the system is doped with holes.
In principle, the maximum additional electron and hole numbers are 6
and 10, respectively.  We calculate for all the possible $\nc$. For
convenience, we call the systems with $|\nc| =1,$ 2 as ``weakly-doped" $\soc$.
We denote the systems with $|\nc| \geq 3$ as ``heavily-doped" $\soc$.
We report the weakly-doped cases separately, because quite interesting changes
are found for these cases.

\subsection{Undoped and weakly-doped C$_{60}$}

First, we show lattice and electronic structures.
Magnitudes of the bond variable
are presented in Table I for $-2 \leq \nc \leq 2$.  Configurations of
the lattice are shown in Fig. 1.
The excess electron density at each site is listed in Table II.
This quantity is calculated by
$[{\sum_{\kappa,s}}^{'} \phi_{\kappa,s}^2 (i)]-1$.
When the system is undoped, there are thirty short bonds and sixty long bonds.
They are shown in Fig. 1(a).
All the sides of the pentagons are the single bonds.
The double and single bonds alternate along
the sides of the hexagons.  This feature is in agreement
with that in the previous study.$^{5,19}$
The lattice configurations of the doped
systems are shown in Fig. 1(b) and (c). The symbols written near bonds,
from a to g, indicate the corresponding bond variables in Table I.
We show three kinds of the shorter bonds.  The
shortest bonds, namely d, are represented by making use of the thick lines.
The second shortest ones, b,
are shown by the usual double lines.  The dashed lines indicate the
third shortest bonds.  They are the bonds f in Fig. 1(b) and bonds g in
Fig. 1(c). Other longer
bonds are not shown for simplicity.    The symbols written
near sites, from A to D, show relations with the electron density in Table II.
The figures are the same for the electron and hole dopings.
When the change in the number of electrons is one,
the change in the electron density is the largest at the sites at the
ends of dashed lines, namely, points D.  The dashed lines are located almost
along an equatorial line of C$_{60}$.  The absolute value of the length of
the bonds g is the smallest of the four kinds of
bonds with negative bond variables.
This implies that the dimerization becomes the weakest along this equatorial
line.  The change in the density is the smallest at the sites at the
ends of the thick lines, points C, when the electron is doped.
It is the smallest at points B, when the hole is doped.
The order of points B and C with respect to the
excess electron density is interchanged.  This is one of the consequence
of the lack of the electron-hole symmetry.
The remarkable feature, i.e., the distortion of the lattice,
is similar to that of a polaron$^{15}$
in polyacetylene; the spatial phase of the alternation of short and long bonds
does not change upon doping, and the lattice distortion
is the largest (the dimerization is the weakest)
where the change in the local electron density is the largest.
Therefore, we conclude that the weakly doped system has lattice structures
and electron distributions like those in a polaron of polyacetylene.
When the change in the electron number is two,
configurations of dashed lines along the equatorial line change, as shown in
Fig. 1(c).  The ordering of bonds, f and g, with respect to
the bond variable is reversed.
Other configurations are the same.  The change in the
electron density is the largest
at points D, too.  Therefore, polaronic distortion persists when the doping
proceeds from one to two electrons (or holes).  This
can be compared with a bipolaron formation in undegenerate conjugated
polymers.$^{15}$  But, we mention this distortion as polaron-like.

Next, we look at changes in the electronic level structures.
Fig. 2(a) shows results of
the undoped and electron-doped systems, while Fig. 2(b) is for
the hole-doped systems.  The line length is proportional to the degeneracy
of a level.  The shortest line is for the undegenerate level.  The arrow
indicates the position of the Fermi level.  It is located between energy
levels when $\nel$ is even.  It is located just at one level if $\nel$
is odd; the energy level at the Fermi level is singly occupied.
In the undoped system, the HOMO is fivefold
degenerate.  The LUMO is threefold degenerate.  This is the consequence
of the lattice geometry.$^5$  Similarly, energy
differences between energy levels remarkably reproduce other results.$^{5,19}$
Therefore, the extended SSH model describes electronic structures
of undoped C$_{60}$ well.  This is due to the fact that
eigenstates around the gap originate mainly from $\pi$-electrons.
We note that there is not the electron-hole symmetry of the SSH model
for polyacetylene.  This is the consequence of the fact that three bonds
are connected to each carbon atom.  Thus, the present system is rather
similar to undegenerate conjugated polymers.

When the system is doped, the degeneracy decreases due to the reduced
symmetry.  This reduction comes from the deformation of the lattice.
The feature of the removal of the degeneracy coincides with that of the
Jahn-Teller theorem.  Therefore, the lattice deformation would be one of the
possible realizations of the Jahn-Teller distortions.  In Ref. 8,
it is predicted that phonon modes with $H_g$ symmetry might couple largely
with electrons in doped $\soc$.  This $H_g$ Jahn-Teller coupling has the
same symmetry as in the present work: the static distortions of the
weakly doped systems are driven by phonons with the $H_g$ symmetry.
The removal of the degeneracies of energy levels has the same splitting
patterns due to the $H_g$ distortion, as shown in Fig. 2 of Ref. 8.
When $|N_{\rm c}| = 1$ and 2, the highest level, which splits
from the HOMO of the undoped system, is undegenerate.
Its energy shifts upward.  In contrast, the other two levels shift only
slightly.  Similarly, the LUMO of the undoped system
splits into two levels.  The energy of the undegenerate level shifts downward,
while change of the energy of the doubly degenerate one is small.  This
change in the level structures is common to two cases of the electron
and hole dopings.  The change
is similar to that in the polaron formation$^{15}$
in polyacetylene; two undegenerate levels split into the gap
from the valence and conduction bands, without change of the gap width.
This feature coincides with the consequence
by the lattice and electron distributions described above.
Since effects of various terms, such as, correlation among electrons,
Coulomb potentials due to charged dopant ions, and possible hopping
interactions between C$_{60}$ molecules are expected to be small,
it is highly possible that the above change of electronic
structures would be observed in experiments, for example, optical absorption.

\subsection{Heavily-doped C$_{60}$}

When the doping proceeds further for larger $|\nc|$, the level structures
change complexly.  When electrons are doped, all the degeneracies are
removed at $\nc = 3$.  Some degeneracies revive at $\nc = 4$ and 5, and the
structure of degeneracies at $\nc = 6$ is like that of the undoped system.
When holes are doped, all the levels are undegenerate at $-\nc = 3, 4, 6,$
and 7.  There are singly and doubly degenerate levels at
$-\nc = 5, 8,$ and 9. Again, the structure of the degeneracies at $-\nc = 10$
is like that of the undoped system.  In Fig. 2, the structures of the
degeneracies show that there is an apparent symmetry between $\nc = 1$
and 5, $\nc = 2$ and 4, $- \nc = 1$ and 9, etc.  We note that those
symmetries should be due to the Jahn-Teller type energy gain.

The lattice structures are very complex when $|\nc| \geq 3$.  For example,
when $\nc=3$, there are forty-five kinds of bond lengths and only
two of all the bands have the same length.
The symmetry of the soccer ball is reduced
strongly.  It is not easy to illustrate lattice patterns.
Rather, it would be more helpful that we shall discuss global features of
changes in lattice patterns.  Therefore, we calculate the mean absolute
values $\langle | y_{i,j} | \rangle$ of the bond variable.
Table III shows the result.  As $|\nc|$ increases, the average
dimerization becomes more and more
weaker.  When six electrons are doped, the average is about 0.03 times of that
of the undoped $\soc$.  The dimerization would be hardly observed
in experiments.  Particularly, lattice fluctuations from the classical
displacements may smear out the very small dimerization even in
low temperatures.  When ten holes are doped, the average becomes about
0.5 times.  In this case, the possibility, that the dimerization would remain
and be observed, is larger than that in the electron-doped case.  In Ref. 6,
the dimerization strength of the electron doped systems is calculated
with the assumption that all of the short bonds change their lengths
by the same value and all of the long bonds do so.  In the present
paper, all of the bond variables are chosen independently.  However,
the dimerization is of the magnitudes similar to that in Ref. 6.

We calculate the total energy gain from that of the undoped system,
and present in Table III.   When electrons are doped, the energy gain
increases as the electron number.  This indicates that electrons are
easily absorbed by the molecule.  In fact, the maximum number $\nc = 6$
has been obtained experimentally.$^{12}$  When holes are doped, the
energy gain saturates at $- \nc = 4$.  It has the maximum at $- \nc
= 6$, and becomes rather smaller at $-\nc = 9, 10$.  This saturation
and reduction of the gain would be related with the more strongly persisting
dimerization in the hole doped systems than in the electron doped ones.
This behavior reflects the lack of the electron-hole symmetry in $\soc$.

\subsection{Optical absorption}

We shall discuss how the ``polarons'' would be observed in experiments.
Particularly, we calculate optical absorption coefficient (dynamical
conductivity).  Our calculations are strongly related with low
energy experiments of radical anions and cations.  Furthermore,
optical absorption of doped lattice systems is interesting, because
the hopping integral between $\soc$ molecules
is very small and negligible in the first
approximation as the recent band calculations$^4$ indicate.

The present calculation is performed with the help of the standard
Kubo formula and information on coordinates of lattice points of
the truncated icosahedron.  The small lattice
displacements are not considered when the system is doped.
This is a valid approximation because
the deformations in the geometry of dipole moments are a higher order effect,
in view of the treatment by the simple extended SSH model where we neglect
a possible coupling of electrons with
phonon modes due to bond-angle modulations and effects of off-plain geometry
around carbon atoms.

Figure 3 shows the results of undoped and electron-doped systems with
$0 \leq \nc \leq 2$, while Fig. 4 is for the hole-doped systems
with $-\nc = 1,$ 2.  We only show data of the weakly-doped cases,
because the general features found
in changes of the absorption patterns persist
when the doping proceeds further.  We adopt the Gaussian broadening procedure
with the width 0.02eV.   Figure 3(a) is the data of the undoped system.
There are two peaks in the figure.  The peak at 2.9eV is the transition between
the HOMO and the next lowest unoccupied molecular orbital.
The other peak at 3.1eV is the transition between the next highest
occupied molecular orbital and the LUMO.  The two transitions are allowed ones.
The transition between the HOMO and the LUMO is forbidden and does not
appear in the figure.  These well-known features$^{4,5}$
are reproduced in the present
calculation.  Figures 3(b) and (c) show the data of the systems with
$\nc = 1$ and 2, respectively.  The two large peaks in Fig. 3(a) now have
small substructures due to the level splittings.  In addition, there appears
a new peak at low energy ($\sim$0.7eV).    This peak corresponds to
the transition between the singly occupied molecular orbital and
the next lowest unoccupied molecular orbital etc., when $\nc = 1$.
It corresponds to the transition between the
LUMO and the next lowest unoccupied
molecular orbital etc., when $\nc = 2$.  Therefore, the new peak at low
energy is the consequence mainly due to the
splitting of the LUMO of the undoped system.
Similar changes are found for hole-doped systems.
Figures 4(a) and (b) show the data for $-\nc = 1$ and 2,
respectively.  The two main peaks in Fig. 3(a)
become broad and have small structures
upon doping.  There appears a new peak.  The energy at about 1eV
is different from that of the electron-doped systems.  This is due to
the lack of the electron-hole symmetry.  The new peak is again the
consequence of the splitting of energy levels by the doping.

\section{UNDOPED AND DOPED C$_{70}$ MOLECULES}

We take the same parameters as in $\soc$,
because effects due to the difference in geometrical
structures are expected to be small.
Number of electrons $N_{\rm el}$ is varied within $N-2 \leq N_{\rm el}
\leq N+2$ ($N = 70$), because weakly-doped cases are particularly
interesting.

Figure 5 shows the geometrical structures in the real space.
Symbols, from a to h, are the names of bonds.  Symbols, from A to E,
are the names of lattice points.  In Table IV, we present the values of
the bond variable $y_{i,j}$.  In Fig. 5, we show only four kinds of shorter
bonds for simplicity.  The double lines with a heavy line indicate the shortest
bonds.  The normal double lines are for the second shortest bonds.
The double lines with dashed and dash-dotted lines depict the
third and fourth shortest bonds, respectively.

First, we discuss results of the undoped $\rug$.  Figure 5(a) shows the
lattice pattern.  The length difference between
the shortest and longest bonds is about 0.057\AA, which agrees well with
the value 0.06\AA{\ } used in Ref. 10.  We also find
that the energy difference between the HOMO and the LUMO, 1.65eV, agrees
with that in Ref. 10.  Therefore, it is shown that the same
parameters used for $\soc$ are realistic for $\rug$.
This means that parameters of the extended SSH model do not sensitively
depend on whether the shape of the molecule resembles a soccer ball or
a rugby ball.  However, there is a structural
difference from Ref. 10.  The length
difference between the bonds, f and g, is about 0.001\AA{\ } and very small.
The dimerization almost disappears at hexagons constructed by bonds, f and g.
These hexagons are similar to benzene rings.  This property coincides with
that of the quantum chemical calculations in Ref. 11.
In Fig. 5(a), the alternation of short and long bonds is present
in the left and right parts of the figure, along bonds from a to e.
It is frequently discussed that $\rug$ is made by cutting of $\soc$
into two parts and adding ten more carbons between them.
The positions of the dimerization, which remains in $\rug$,
correspond to those in $\soc$.  Table V shows the electron density.
The electron distribution is not uniform differently from the undoped $\soc$.
Spatial oscillation, from the site A to E, is seen.  This might be caused
by the distorted structure of $\rug$.

Next, we look at changes in lattice structures and electron distributions
of the doped $\rug$.  Figure 5(b) shows the lattice configuration when one
electron is added or removed.  Figure 5(c) depicts the case where
two electrons are added or removed.  Figures are common to electron and
hole dopings.  When one electron is added to or removed from the undoped
$\rug$, the two kinds of bonds with heavy and normal lines are interchanged.
Positions of the other bonds shown by dashed and dash-dotted lines
do not change.  When one electron is added
or removed further, bonds with dash-dotted lines change their positions.
Other shorter bonds do not interchanged.  Through this development,
the patterns with the mirror reflection symmetry are retained.  This
is the consequence of the shape like a rugby ball.  In Table VI, we list the
excess electron density, where the electron density of the undoped system
is subtracted.   Change in electron density at sites E
is very small.  This is the consequence
of the fact that dimerization almost
disappears along bonds, f and g, in the undoped $\soc$.  The part
along the equatorial line has the property similar
to that of the graphite plane.  The strengths of the dimerization
change largely along bonds, from a to e, upon doping.  The additional
charges favor to accumulate near these bonds.  Therefore,
the density of the additional charge would be very small at sites E.
The positions D, where the additional charges accumulate most densely,
correspond to the sites D of $\soc$ in Fig. 1.
These sites lie along the equatorial
line of $\soc$.  When we make $\rug$
from $\soc$, sites E are added in the interval.
But, the property, that additional charges favor to accumulate
at sites D, persists for $\rug$.  We believe that
this finding is quite interesting.

Finally, we show structures of electronic energy levels in Fig. 6.
As in $\soc$, electronic states around the LUMO and HOMO
are originated mainly from $\pi$-electrons.  Therefore, results are
reliable as far as we look at levels only around the Fermi level.  This energy
region is particularly interesting in optical absorption experiments.
In Fig. 6, energy levels in the region from -3eV to 1.5eV are shown
varying the electron number.  Notations are the same as those in Fig. 2.
When the system is half-filled, the HOMO and LUMO are undegenerate.
The energy difference between them is 1.65eV, which is comparable
to that of $\soc$ (2.26eV).  Positions and degeneracies of
other levels below the HOMO and above the LUMO show
an overall agreement with the other result.$^{10}$
This clearly indicates the validity of the extended SSH model also.
The energy difference between the HOMO and the next highest occupied level
is much smaller than the width of the gap.  Similarly, the energy difference
between the LUMO and the next lowest unoccupied level is small.
This is the consequence of the smallness of the structural perturbation from
$\soc$ to $\rug$.  The structure of the degeneracies (the property that there
are onefold and twofold degenerate levels) is like that
in $\soc$ where one or two electrons (or holes) are doped.  This
is due to the same symmetry of those systems.
When the system is doped with up to two electrons
or holes, a significant change is predicted.   The HOMO and LUMO of the
undoped system extend into the gap apparently.  The positions of the
other levels change only slightly.  This property has been found also
in doped $\soc$.  We have named this change as the ``polaron-like change"
in polyacetylene.  The magnitude of level intrusion is larger than that in
$\soc$.  This is the consequence of the fact that levels near the gap have
already split at $\nc = 0$.
The HOMO and LUMO of the undoped system have large
amplitude at sites, from A to D.  The amplitude at D is the largest.
The amplitude at E is very small.  Therefore, the additional charge
favor to accumulate most at sites D.

\section{CONCLUDING REMARKS}

We have applied the extended SSH model to undoped and doped
$\soc$ and $\rug$.  The calculations of the undoped system ($\soc$ or $\rug$)
agree well with the known results.$^{5,10,11}$
When the system is doped with one or two
electrons (or holes), the additional charges accumulate almost
along an equatorial line of the molecule.
The dimerization becomes weaker almost along the same line.
In the energy level structure, two levels intrude largely in the gap.
These changes are characteristic to ``polaron excitations'' in
conjugated polymers.$^{15}$  We have discussed that the changes can be observed
by experiments, for example, optical absorption, and have calculated
dynamical conductivity for $\soc$ in order to look at how they would be
observed.  It is predicted that there appears a new peak at the
lower energy than the intergap transition peaks.

We have found that sites where additional charges
favor to accumulate are common to $\soc$ and $\rug$.  They are
along the equatorial line in $\soc$.  In $\rug$, ten more atoms are inserted
among these sites.  Wave functions
of the intragap levels have large amplitudes at corresponding sites, too.
The additional charges occupy these wave functions, and thus the
sites, where additional charge density is larger, are common.
This should be compared to the statement that $\rug$ is made from
$\soc$, by cutting into two parts and adding ten more carbons.
It is quite interesting that there are relations with respect to
electronical properties as well as the
above-mentioned structural relation between (doped as well as undoped)
$\soc$ and $\rug$.

We have not considered various effects, because the study is at the first stage
of the investigation.  Electron correlations among electrons
might effect on energy level structures, particularly when the
Fermi level lies just at one energy level.  Fluctuation
of phonons from the classical mean field might effect on the
stability of the polaronic distortions.  Interactions between molecules
would be a necessary problem also: it is an important problem to clarify
how our findings persist especially in doped crystals.  These neglected
contributions pose interesting problems for future investigations.

Recently, after the submission and
the acceptance of the previous paper,$^{16}$
we have found that Friedman$^{21}$ has independently
performed the same calculation of electron-doped $\soc$.
The result is common as far as the comparison is made.

\begin{flushleft}
{\bf ACKNOWLEDGEMENTS}
\end{flushleft}

Fruitful discussions with Drs. K. Yamaji, S. Abe, Y. Asai
(Electrotechnical Laboratory), A. Terai (University of Tokyo),
N. Hamada, S. Saito, and A. Oshiyama (Fundamental Research
Laboratories, NEC Corporation) are acknowledged.
The author thanks Prof. B. Friedman (Sam Houston State University)
for sending his preprint prior to publication.

\pagebreak
\begin{flushleft}
{\bf REFERENCES}
\end{flushleft}

\noindent
$^*$Electronic mail address: e9118@etlcom1.etl.go.jp, e9118@jpnaist.bitnet\\
$^1$A.F. Hebard, M.J. Rosseinsky, R.C. Haddon, D.W. Murphy,
S.H. Glarum, T.T.M. Palstra, A.P. Ramirez, and A.R. Kortan,
Nature {\bf 350}, 600 (1991).\\
$^2$M.J. Rosseinsky, A.P. Ramirez, S.H. Glarum, D.W. Murphy,
R.C. Haddon, A.F. Hebard, T.T.M. Palstra, A.R. Kortan, S.M. Zahurak,
and A.V. Makhija, Phys. Rev. Lett. {\bf 66}, 2830 (1991).\\
$^3$K. Tanigaki, T.W. Ebbesen, S. Saito, J. Mizuki, J.S. Tsai,
Y. Kubo, and S. Kuroshima, Nature {\bf 352}, 222 (1991).\\
$^4$S. Saito and A. Oshiyama, Phys. Rev. Lett. {\bf 66}, 2637 (1991).\\
$^5$M. Ozaki and A. Takahashi, Chem. Phys. Lett. {\bf 127}, 242 (1986);
S. Saito, in {\sl Clusters and Cluster Assembled Materials}, eds.
R. S. Averback, D. L. Nelson, and J. Bernholc
(Materials Research Society, Pittsburgh, 1991).\\
$^6$F. C. Zhang, M. Ogata, and T.M. Rice, Phys. Rev. Lett. {\bf 67},
3452 (1991).\\
$^7$Y. Asai, (preprint)\\
$^8$C.M. Varma, J. Zaanen, and K. Raghavachari, Science {\bf 254},
989 (1991).\\
$^9$M.J. Rice, H. Y. Choi, and Y.R. Wang, Phys. Rev. B {\bf 44},
10414 (1991).\\
$^{10}$S. Saito and A. Oshiyama, Phys. Rev. B {\bf 44}, 11532 (1991).\\
$^{11}$G.E. Scuseria, Chem. Phys. Lett. {\bf 180}, 451 (1991);
J. Baker, P.W. Fowler, P. Lazzeretti, M. Malagoli, and R. Zanasi,
Chem. Phys. Lett. {\bf 184}, 182 (1991).\\
$^{12}$T. Takahashi, T. Morikawa, S. Sato, H. Katayama-Yoshida,
A. Yuyama, K. Seki, H. Fujimoto, S. Hino, S. Hasegawa, K. Kamiya,
H. Inokuchi, K. Kikuchi, S. Suzuki, K. Ikemoto, and Y. Achiba, Physica
C {\bf 185-189}, 417 (1991); C.T. Chen, L.H. Tjeng, P. Rudolf, G. Meigs,
L.E. Rowe, J. Chen, J.P. McCauley Jr., A.B. Smith III, A.R. McGhie,
W.J. Romanow, and E.W.Plummer, Nature {\bf 352}, 603 (1991).\\
$^{13}$T. Kato, T. Kodama, M. Oyama, S. Okazaki, T. Shida, T. Nakagawa,
Y. Matsui, S. Suzuki, H. Shiromaru, K. Yamauchi, and Y. Achiba,
Chem. Phys. Lett. {\bf 180}, 446 (1991).\\
$^{14}$W.-P. Su, J.R. Schrieffer, and A.J. Heeger, Phys. Rev. B {\bf 22},
2099 (1980).\\
$^{15}$ A.J. Heeger, S. Kivelson, J.R. Schrieffer, and W.-P. Su,
Rev. Mod. Phys. {\bf 60}, 781 (1988).\\
$^{16}$K. Harigaya, J. Phys. Soc. Jpn. {\bf 60}, 4001 (1991).\\
$^{17}$K. Harigaya, Chem. Phys. Lett., (to be published).\\
$^{18}$K. Harigaya, A. Terai, Y. Wada, and K. Fesser, Phys. Rev. B
{\bf 43}, 4141 (1991).\\
$^{19}$G.W. Hayden and E.J. Mele, Phys. Rev. B {\bf 36}, 5010 (1987).\\
$^{20}$C.S. Yannoni, P.P. Bernier, D.S. Bethune, G. Meijer, and J.R. Salem,
J. Am. Chem. Soc. {\bf 113}, 3190 (1991).\\
$^{21}$B. Friedman, Phys. Rev. B {\bf 45}, 1454 (1992).\\

\pagebreak
\begin{flushleft}
{\bf TABLE CAPTIONS}
\end{flushleft}

\noindent
TABLE I.  The bond variable for the undoped and doped $\soc$ with
$-2 \leq \nc \leq 2$.  The bond length becomes longer from the top
to the bottom.  The unit is \AA.  We show
the symbols which indicate the positions in Fig. 1 and
the number of bonds of the same length, in the square brackets.

{}~

\begin{tabular}{rrrrr} \hline
     $\nc=0$ &             1&             2&            -1&            -2 \\
\hline
 0.03333[30] & 0.03272[d,10]& 0.03236[d,10]& 0.03808[d,10]& 0.04316[d,10] \\
-0.01667[60] & 0.02945[b,10]& 0.02578[b,10]& 0.03447[b,10]& 0.03565[b,10] \\
             & 0.02032[f,10]& 0.01056[g,10]& 0.02099[f,10]& 0.01428[g,10] \\
             &-0.00328[g,10]& 0.00696[f,10]&-0.00123[g,10]& 0.00961[f,10] \\
             &-0.01442[a,10]&-0.01222[a,10]&-0.01595[a,10]&-0.01519[a,10] \\
             &-0.01497[c,20]&-0.01299[c,20]&-0.01596[c,20]&-0.01520[c,20] \\
             &-0.01742[e,20]&-0.01873[e,20]&-0.02221[e,20]&-0.02805[e,20] \\
\hline
\end{tabular}

{}~~~~~~~~~~~~~~~~~~~~~~~~~

\noindent
TABLE II. The excess electron density per site of doped $\soc$.
The symbols indicate the positions of lattice sites shown in Fig. 1.

{}~

\begin{tabular}{crrrr} \hline
Position & $\nc=1$ &       2 &      -1 &      -2 \\ \hline
A        & 0.00475 & 0.00923 &-0.00108 &-0.00214 \\
B        & 0.02226 & 0.04249 &-0.00059 &-0.00122 \\
C        & 0.00145 & 0.00334 &-0.01100 &-0.02267 \\
D        & 0.03504 & 0.07080 &-0.03817 &-0.07565 \\ \hline
\end{tabular}

\newpage

\noindent
TABLE III.  The mean absolute values of the bond variable,
$\langle | y_{i,j} | \rangle$, for the undoped
and doped $\soc$ with $-10 \leq \nc \leq 6$.  The total energy gain
from that of the undoped system is shown, also.

{}~

\begin{tabular}{rrc} \hline
$\nc$ (\AA) & $ \langle |y_{i,j}| \rangle $ & Energy gain (eV) \\ \hline
0 & 0.02222 & $\cdot \cdot \cdot$ \\ \hline
1 & 0.01833 & 0.0626 \\
2 & 0.01681 & 0.2542 \\
3 & 0.01242 & 0.3333 \\
4 & 0.01035 & 0.5386 \\
5 & 0.00506 & 0.6331 \\
6 & 0.00061 & 0.8516 \\ \hline
-1 & 0.02079 & 0.0852 \\
-2 & 0.02260 & 0.3420 \\
-3 & 0.02079 & 0.3567 \\
-4 & 0.02103 & 0.5180 \\
-5 & 0.01883 & 0.4894 \\
-6 & 0.01711 & 0.5713 \\
-7 & 0.01491 & 0.4715 \\
-8 & 0.01381 & 0.5123 \\
-9 & 0.01082 & 0.3211 \\
-10& 0.00973 & 0.2930 \\ \hline
\end{tabular}

\newpage

\noindent
TABLE IV.  The bond variable for undoped and doped $\rug$.
The bond length becomes longer from the top to the bottom.
The unit is \AA.  We show
the symbols which indicate the positions in Fig. 5 and
the number of bonds of the same length, in the square brackets.

{}~

\begin{tabular}{rrrrr} \hline
       $\nc=0$ &             1&             2&            -1&            -2 \\
\hline
 0.03514[d,10] & 0.03126[b,10]& 0.03146[b,10]& 0.03721[b,10]& 0.04346[b,10] \\
 0.03114[b,10] & 0.02481[d,10]& 0.01419[d,10]& 0.03200[d,10]& 0.02876[d,10] \\
 0.01109[g,20] & 0.01010[g,20]& 0.00906[g,20]& 0.01122[g,20]& 0.01133[g,20] \\
 0.01030[f,10] &-0.00075[f,10]& 0.00150[e,20]&-0.00031[f,10]&-0.00487[e,20] \\
-0.01532[c,20] &-0.01047[e,20]&-0.01104[h,~5]&-0.01321[h,~5]&-0.01089[f,10] \\
-0.01535[h,~5] &-0.01329[h,~5]&-0.01212[f,10]&-0.01356[e,20]&-0.01094[h,~5] \\
-0.01606[a,10] &-0.01571[c,20]&-0.01611[c,20]&-0.01911[c,20]&-0.02284[a,10] \\
-0.02220[e,20] &-0.01651[a,10]&-0.01681[a,10]&-0.01940[a,10]&-0.02297[c,20] \\
\hline
\end{tabular}
\newpage

\noindent
TABLE V. The electron density per site of the undoped $\rug$.
The symbols indicate positions of lattice sites shown in Fig. 5.

{}~~~~~~~~~~~~~~~~~~~~~~~~~~

\begin{tabular}{cr} \hline
Position & $\nc=0$ \\ \hline
A        & 0.99199 \\
B        & 1.01246 \\
C        & 0.99540 \\
D        & 1.01151 \\
E        & 0.98174 \\ \hline
\end{tabular}

{}~

\noindent
TABLE VI.  The excess electron density per site of the doped $\rug$.
The symbols indicate positions of lattice sites in Fig. 5.

{}~

\begin{tabular}{crrrr} \hline
Position & $\nc=1$ &       2 &       -1 &       -2 \\ \hline
A        &-0.00362 &-0.00705 & -0.01256 & -0.02504 \\
B        & 0.00215 & 0.00442 & -0.01318 & -0.02708 \\
C        & 0.02499 & 0.04497 & -0.00714 & -0.01443 \\
D        & 0.02610 & 0.05203 & -0.03016 & -0.05989 \\
E        &-0.00072 &-0.00138 &  0.00034 &  0.00075 \\ \hline
\end{tabular}

\pagebreak
\begin{flushleft}
{\bf FIGURE CAPTIONS}
\end{flushleft}

\noindent
FIG. 1. Lattice configurations of
(a) the undoped $\soc$, (b) the doped
$\soc$ with $|\nc|=1$, and (c) the doped $\soc$ with $|\nc| =2$.
Figures are common to electron and hole dopings. Symbols indicate
names of bonds and lattice sites.
In (a), short bonds are shown by the double lines.
In (b) and (c), three kinds of the shorter bonds are shown.  The thick lines
indicate the shortest bonds, while the dashed ones are for the
third shortest bonds.

{}~

\noindent
FIG. 2. Energy level structures of (a) the undoped and electron-doped
$\soc$, and (b) hole-doped $\soc$.
The line length is proportional to the degeneracy
of the energy level.  The shortest line is for the  undegenerate levels.
The arrow indicates the position of the Fermi level.

{}~

\noindent
FIG. 3.  Optical absorption of the undoped and electron-doped $\soc$.
The unit of the ordinate is arbitrary.  In (a), the result of the undoped
$\soc$ is shown.  Figures (b) and (c) are the data of the systems
with $\nc = 1$ and 2, respectively.

{}~

\noindent
FIG. 4. Optical absorption of the hole-doped $\soc$.
The unit of the ordinate is arbitrary.  Figures (a) and (b)
are the data of the systems with $-\nc = 1$ and 2, respectively.

{}~

\noindent
FIG. 5. Lattice configurations for (a) the undoped $\rug$, (b) the
doped $\rug$ with $|\nc|=1$, and (c) the doped $\rug$ with $|\nc| =2$.
Figures are common to electron and hole dopings.  Symbols are
names of bonds and lattice sites.
The double lines with a heavy line indicate the shortest
bonds.  The normal double lines are for the second shortest bonds.
The double lines with dashed and dash-dotted lines depict the
third and fourth shortest lines, respectively.

{}~

\noindent
FIG. 6. Energy level structures around the HOMO and LUMO of the undoped
and doped $\rug$.  The line length is proportional to the degeneracy
of the energy level.  The shortest line is for the  undegenerate levels.
The arrow indicates the position of the Fermi level.


\end{document}